\documentclass{aa}
\usepackage{epsfig,graphicx}

\newcommand{\rxx}{\mbox{RX\,J0925.7$-$4758}}
\newcommand{\rx}{\mbox{MR~Vel}}
\newcommand{\cmsq}{{\rm cm}^{-2}}
\newcommand{\ergs}{{\rm erg\,s}^{-1}}
\newcommand{\gtap}{\mathrel{\hbox{\rlap{\lower.55ex \hbox {$\sim$}}
                   \kern-.3em \raise.4ex \hbox{$>$}}}}
\newcommand{\ltap}{\mathrel{\hbox{\rlap{\lower.55ex \hbox {$\sim$}}
                   \kern-.3em \raise.4ex \hbox{$<$}}}}
\newcommand{\msun}{M_{\odot}}
\newcommand{\msunyr}{M_{\odot}{\rm \,yr}^{-1}}

\begin{document}

\title{The complex X-ray spectrum of the supersoft source \rx}
\titlerunning{X-ray spectrum of \rx}

\author{Henk Bearda\inst{1} 
\and Wouter Hartmann\inst{2} 
\and Ken Ebisawa\inst{3} 
\and John Heise\inst{1,2} 
\and Jelle Kaastra\inst{2}
\and Rob van der Meer\inst{2} 
\and Frank Verbunt\inst{1} 
\and Christian Motch\inst{4}
}
\authorrunning{Henk Bearda et al.}

\offprints{Henk Bearda}
\mail{bearda@astro.uu.nl}

\institute{   Astronomical Institute, Utrecht University,
              P.O.Box 80000, NL-3508 TA Utrecht, The Netherlands
         \and SRON National Institute for Space Research,
              Sorbonnelaan 2, NL-CA 3584 Utrecht, The Netherlands,
         \and Laboratory for High Energy Astrophysics, NASA 
              Goddard Space Flight Center, Greenbelt, MD 20771,
              U.S.A.
         \and Observatoire Astronomique, UA 1280 CNRS,
              11 rue de l'Universit\'e, F-67000 Strasbourg,
              France
              }

\date{Received date \today / Accepted date}   

\abstract{We present the X-ray spectrum of \rx\,/\,\rxx, obtained with the
Medium Energy Grating spectrometer of the Chandra X-ray Telescope.
The simplest models used by earlier authors, stellar atmospheres in
combination with a thermal plasma in collisional ionization equilibrium, 
cannot explain the spectrum. Neither does a photo-ionized plasma.
We identify P\,Cygni profiles of Fe\,XVII and O\,VIII, from which we conclude
that these lines arise in a wind.
We conclude that major uncertainty exists about the bolometric
luminosity of \rx, and perhaps of supersoft sources in general,
so that the theoretical prediction that this luminosity 
derives from steady nuclear burning cannot be verified.
}

\maketitle

\section{Introduction}

Supersoft X-ray sources are called such because they emit most of
their X-ray flux at energies less than 0.4\,keV.
First discovered with the Einstein satellite (Long et al.\ 1981), 
they were recognized as a class only after the more sensitive 
observations with ROSAT (Tr\"umper et al.\ 1991).
The suggestion that the X-ray emission of supersoft sources is caused
by steady nuclear burning at the surface of white dwarfs with masses
near the Chandrasekhar limit (Van den Heuvel et al.\  1992, see
also Shara et al.\ 1977, Iben 1982) gained credibility when 
the nova GQ Mus was found to be a supersoft source (\"Ogelman et al.\ 1993).
\nocite{lhg81}\nocite{tha+91}
\nocite{hbnr92}\nocite{iben82}\nocite{ooks93}\nocite{sps77}

Van den Heuvel et al.\ (1992) argued that the high mass-transfer rates 
($\gtap\,10^{-7}\,\msunyr$)
demanded by theory for steady-state hydrogen burning require
donors with masses in the range of 1.4-2.2\,$\msun$.
However, optical identifications soon showed that the
class of supersoft X-ray sources is far from homogeneous, in
some cases exclude $\gtap\,1.4\,\msun$ donors, and also
cast doubt on the premise that the white dwarfs in these systems
accrete steadily at high rates (e.g.\ the review by G\"ansicke et al.\ 2000).
\nocite{gtbr00}

Black body fits indicated that the X-ray luminosities of supersoft
sources -- several of which are located in the Magellanic Clouds, and
thus have known distances -- are in excess of the
Eddington limit for a white dwarf (Greiner et al.\ 1991).
Fitting LTE white-dwarf atmosphere model
spectra to the X-ray data, Heise et al.\ (1994)
showed that the X-ray luminosities may well be at or below the
Eddington limit. NLTE models may give very different
luminosities again (Hartmann \&\ Heise 1997).
The rather limited spectral resolution of the Einstein and
ROSAT X-ray data prevents an accurate determination of the
X-ray spectrum and of the
interstellar absorption, and this leads to very large
uncertainty of the estimated bolometric flux.
Accurate determination of the luminosities is important
for comparison with the models, and to establish the
relation of the supersoft sources
with other sources showing very soft spectra at low luminosities, such
as the variable source in the globular cluster NGC\,5272 
whose nature is unknown
(Dotani et al.\ 1999) or V\,Sge 
whose X-rays may arise from colliding winds
(e.g.\ Wood \&\ Lockley 2000).
\nocite{ghk91}\nocite{htk94}\nocite{hh97}\nocite{dag99}\nocite{wl00}

Studies of Cal\,87 and \rx\,/\,\rxx, both supersoft sources 
with sig\-ni\-fi\-cant count rates above 0.4\,keV, have shown
that an LTE or NLTE white-dwarf atmosphere model does not correctly
describe the X-ray spectrum (Hartmann et al.\ 1999, Ebisawa et al.\ 2001).
In this paper we present the first X-ray spectrum of the
supersoft source \rx\ with sufficient resolution to test 
spectral models in some detail.
\nocite{hhk+99}\nocite{emk+01}

\vfil
~

\begin{table*}
\caption{Summary of models for the X-ray data of \rx, giving
the instrument used, its wavelength range and the resolution at 1\,keV, the model
used and its fit parameters. BB, SA and TP indicate a black body, 
NLTE stellar atmosphere model, and thermal plasma in collisional ionization
equilibrium,
respectively. An index 2 indicates the second spectral component.
Luminosities $L$ are bolometric for an assumed distance of 1\,kpc; 
column densities $N_{\rm H}$ are the same for both components, where
applicable. 
References a.\ Motch et al.\ 1994 b.\ Hartmann et al.\ 1999 c.\ Ebisawa et al.\ 2001
d.\ this paper
\label{tabmod}}
{\resizebox{\textwidth}{!}{\begin{tabular}{lllllllrllrll}
  &instrument   &E        &$\Delta E/E$&model &$T$         &$g$          &$L$            &$T_2$       &$L_2$           &$N_{\rm H}$       &$\chi^2_r$/dof \\
  &             &(keV)    &(1 keV)     &      &(10$^5$\,K) &(cm\,s$^{-2}$)&($\ergs$)      &(10$^5$\,K) &($\ergs$)      &($\cmsq)$         &               \\
a &ROSAT PSPC   &0.1-2.4 &0.4         &BB    &5.6         &  --         &$5\,10^{38}$   &            &                &$1.7\,10^{22}$    &1.1/17         \\
b &SAX LECS     &0.1-10. &0.15        &SA    &8.75(5)     &$10^9$       &$4.1\,10^{36}$ &            &                &$1.5(2)\,10^{22}$ &5.6/15         \\
b &SAX LECS     &0.1-10. &0.15        &SA+TP &8.75(4)     &$10^9$ &$2.9\,10^{36}$ &80(60)      &$<\,1.5\,10^{35}$ &$1.4(2)\,10^{22}$ &3.4/14         \\
c &ASCA SIS     &0.4-10. &0.08        &BB    &5.2         &--           &$1.8\,10^{40}$ &            &                &$2.1\,10^{22}$    &8.1/73         \\
c &ASCA SIS     &0.4-10. &0.08        &SA+TP &10.9        &$10^{10}$    &$7\,10^{35}$   &12          &                &$1.3\,10^{22}$    &2.0/71         \\
d &Chandra HETG &0.4-8.0 &0.02        &BB    &5.2         & --          &$2.8\,10^{39}$ &            &                &$1.6\,10^{22}$    &2.5/252        \\
d &Chandra HETG &0.4-8.0 &0.02        &SA    &8.97        &$10^9$       &$5.3\,10^{35}$ &            &                &$1.0\,10^{22}$    &2.9/252        \\
d &Chandra HETG &0.4-8.0 &0.02        &SA+TP &8.96        &$10^9$       &$4.1\,10^{35}$ &20.6        &$4.9\,10^{34}$  &$1.0\,10^{22}$    &2.6/250        \\
 & \\
\end{tabular}}}
\end{table*}

\section{Earlier observations of \rx}

\rx\ was discovered in the Galactic Plane in the ROSAT All Sky Survey, 
in a search for
supersoft sources. The intrinsic softness of the source
combined with strong interstellar absorption cause most of the photons
of this source to be detected between 0.5-1.5\,keV. 
A black body fit to the data from a ROSAT PSPC pointing
gives a temperature of $\sim\,48$\,eV, for an interstellar absorption
column of $\sim\,1.7\,10^{22}\,\cmsq$ (Motch et al.\ 1994).  Observations
with the BeppoSAX MECS and the ASCA SIS show excess flux at energies
$\gtap\,1.2$\,keV with respect to the flux predicted by black body or
stellar atmosphere models. To explain this, a thermal
plasma has been suggested as a second component of the spectrum.
However, models combining a NLTE white-dwarf atmospheric model with
a thermal plasma in collisional ionization equilibrium do not provide 
acceptable fits to the data, which appear to show absorption edges 
at 0.87\,keV (O\,VIII), 1.02\,keV (unidentified), and 1.36\,keV (Ne\,X),
according to Ebisawa et al.\ (2001).
\nocite{mhp94}

NLTE atmosphere models give
a much lower (unabsorbed) flux for the observations than LTE models, 
which in turn give lower fluxes than black body fits.
Whereas LTE models can give near-Eddington luminosities for \rx\ at
a distance of 1\,kpc, NLTE models lead to much larger distances
or lower luminosities.

Spectral fits to pre-Chandra X-ray data are summarized in Table\,\ref{tabmod}.

A 17$^{th}$ magnitude optical counterpart was identified on the basis of a 
ROSAT HRI position (Motch et al.\ 1994) 
and showed a photometric and spectroscopic period of 4.029\,d 
(Schmidtke et al.\ 2001). The optical lightcurve
is asymmetric and single-peaked, with a full amplitude of about
0.3\,mag. A shorter, unexplained period of about 0.25\,d also appears
to be present in the photometry.
The very red optical continuum -- presumably of the accretion disk --
and the very strong interstellar absorption lines
indicate a high interstellar absorption, in agreement with the
X-ray results (Motch et al.\ 1994). Transient satellite lines
of the H\,$\alpha$ line indicate a transient double-sided jet,
with a velocity of $\sim\,5000$\,km\,s$^{-1}$, and an opening angle
between 34 and 80\,degrees (Motch 1998).
The photometry shows  no eclipses, indicating an inclination
$i\ltap\,65^{\circ}$.
\nocite{mot98}\nocite{sc01}
 
The radial velocity curve of the He\,II\,4686 and H\,$\alpha$ lines,
with a semi-amplitude of 75\,km\,s$^{-1}$, with the absence of eclipses
lead to a likely donor mass between 1.0 and 1.6\,$\msun$ for assumed masses
between 0.5 and 1.4\,$\msun$ for the white dwarf.
The absence of the donor in the visual spectrum indicates a
bright disk, which in turn indicates a minimum distance of a kpc.

\section{Chandra observations and data analysis}

\rx\ was observed on 14/15 November 2000, for a net exposure time
of 56.5\,ks,
using the High-Energy Transmission Grating Spectrometer (HETGS) of the
Chandra X-ray Observatory (Canizares et al.\ 2000 and references
therein). The high energy grating (HEG) could not be used for analysis
because of a low S/N-ratio. 
The medium energy grating (MEG) is optimized for the range of 1.2-30\,\AA\ 
and provides a spectral resolution of 0.02\,\AA\ FWHM.
The photons are collected at an array of CCDs, ACIS-S, in the focal plane.
\nocite{chd+00}

The data were processed with standard Chandra X-ray Center software
CIAO-2.1, as follows. Photons are extracted by applying a spatial
filter around the spectral image, and negatively and positively
dispersed spectra are added. Most background photons are recognized by
pulse height analysis, and removed; the higher-order spectra of this
faint source do not contain useful information.  
Flux calibration was done with the 1999-07-22 version
of the detector response matrix, part of the Chandra calibration 
database CALDB\,2.3.

\begin{figure*}[t]
\epsfig{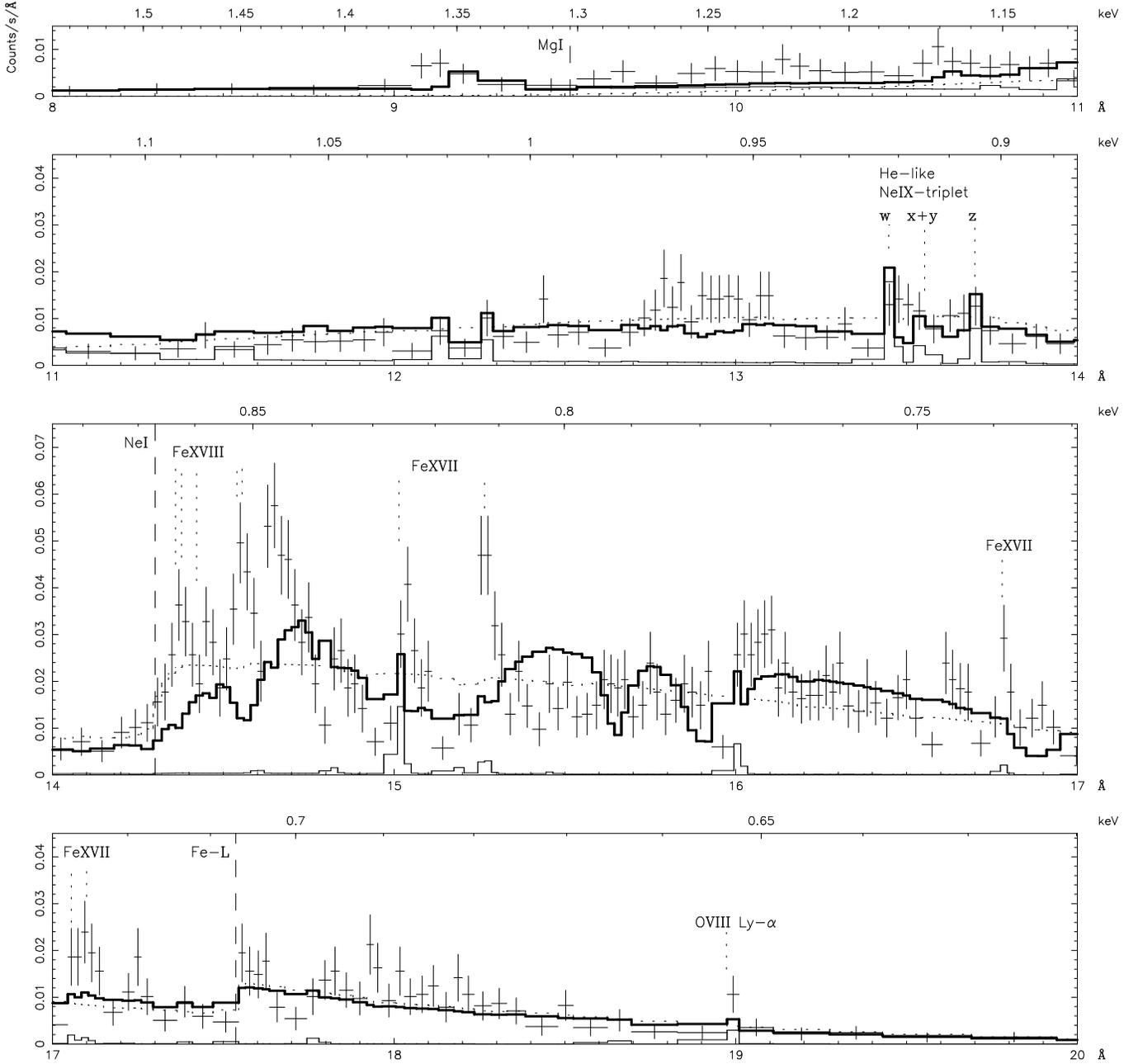}
\caption{The observed X-ray spectrum of \rx\, binned to the minimum width
required for either 1/3 of the instrumental spectral resolution or
20 counts, together with the best black body fit (dotted line)
and the best fit combination (thick line) of the spectra from a 
NLTE atmosphere and a thermal plasma in collisional ionization equilibrium
(the latter is also shown separately with a thin line). 
Some identified spectral features are indicated.
\label{spectrum}}
\end{figure*}

The errors in each bin with $N$ photons are approximately Poissonian
and given by (see Gehrels 1986):
\begin{equation}
\Delta N = 1+\sqrt{N+0.75}
\end{equation}
For comparison with model spectra, we use 
$\chi^2$ fitting and bin the data, until the oversampling
of the spectral resolution is less than a factor 3, and until 
at least 20 counts are present in each energy bin.
The resulting spectrum is shown in Fig.\,\ref{spectrum}.
\nocite{geh86}

\section{Model spectra}

\begin{table}
\centerline{\begin{tabular}{|l|rl|c|rl|}
\hline
\hbox to0.20\columnwidth{ion\hfil}&
	\multicolumn{2}{|c|}{\hbox to0.20\columnwidth{\hfil number of\hfil}}&
			\hbox to0.20\columnwidth{\hfil quantum\hfil}&
				\multicolumn{2}{|c|}{\hbox to0.20\columnwidth{\hfil number of\hfil}}\\
&
	\multicolumn{2}{|c|}{levels}&
			number&
				\multicolumn{2}{|c|}{transitions}\\
\hline
H\,\textsc{I}           
	&\hbox to0.12\columnwidth{\hfil 5} 
		&
			&5     
				&\hbox to0.14\columnwidth{\hfil --}     
					&\\
He\,\textsc{II}         &3      &&3     &--     &\\
C\,\textsc{V}           &17     &&4     &315    &\\
C\,\textsc{VI}          &10     &&4     &285    &\\
N\,\textsc{VI}          &17     &&4     &315    &\\
N\,\textsc{VII}         &10     &&4     &285    &\\
O\,\textsc{VI}          &5      &&3     &153    &\\
O\,\textsc{VII}         &17     &&4     &315    &\\
O\,\textsc{VIII}        &10     &&4     &285    &\\
Ne\,\textsc{VII}        &6      &&2     &2111   &\\
Ne\,\textsc{VIII}       &5      &&3     &153    &\\
Ne\,\textsc{IX}         &17     &&4     &315    &\\
Ne\,\textsc{X}          &10     &&4     &285    &\\
Fe\,\textsc{XV}         &19     &&3     &5074   &\\
Fe\,\textsc{XVI}        &7      &&4     &184    &\\
Fe\,\textsc{XVII}       &21     &&3     &3409   &\\
Fe\,\textsc{XVIII}      &2      &&2     &9437   &\\
Fe\,\textsc{XIX}        &6      &&2     &23915  &\\
Fe\,\textsc{XX}         &8      &&2     &37795  &\\
Fe\,\textsc{XXI}        &12     &&2     &34017  &\\
\hline
\end{tabular}}
\caption{The ions used for the construction of the
model atmospheres in TLUSTY, with the respective 
number of energy levels and the corresponding quantum number up to 
which all levels are considered.
The number of line transitions used by SYNSPEC are also given; 
these are all transitions between energy levels with quantum
numbers from 1 to 10.}
\label{ions}
\end{table}

We compute NLTE model spectra for white dwarf atmospheres,
characterized by effective temperature, gravitational acceleration,
and elemental abundances, as described in Hartmann et al. (1999),
i.e.\ using the codes TLUSTY (version 195) for the construction of a model
atmosphere, and SYNSPEC (version 42) for the computation of the 
emergent spectrum.  
These codes were developed and are described by Huben\'y (1988)
and Huben\'y \&\ Lanz (1995).
For the atmosphere model we use 70 depth zones, and a 
frequency range from $10^{12}$\,Hz to $9.6\,10^{17}$\,Hz;
convergence is considered to have been achieved when the
maximum relative change between subsequent model parameters is less 
than 0.001.
The ions and transitions used in our computations are listed
in Table\,\ref{ions}; their numbers are limited by
computational expense.
The atomic data are extracted from the Opacity Project with 
TOPbase (see Cunto et al.\ 1993).
\nocite{hub88}\nocite{hl95}\nocite{cmo+93}

Models for thermal plasmas in collisional ionization equilibrium 
were computed in SPEX, with the code developed and described by
Kaastra et al.\ (1996).

\vfil
~

\section{Results}

The observed spectrum shown in Fig.\,\ref{spectrum} shows the 
following emission
lines (with theoretical wavelengths from Phillips et al.\ 1999).
The lines at $\lambda$\,15.014, 15.265, 16.780, 17.055, 
and 17.100\,\AA\ are all due to Fe\,XVII.
The line at $\lambda$\,18.973\,\AA\ is the O\,VIII Lyman\,$\alpha$ line.
The strong line complex between $\lambda$\,14.3-14.6\,\AA\ is a blend of
Fe\,XVIII lines, which cannot be identified individually, the strong
feature at $\lambda$\,14.6-14.7\,\AA\ may be a blend of O\,VIII 
and Fe\,XIX lines. 
The complex of the Ne\,IX $\lambda$\,13.448, 13.553 and 13.700\AA\ 
triplet, respectively the resonance line (w), the intercombination
(x+y) lines, and the forbidden  (z) line of the He-like ion; these
lines are not recognizable individually, but their presence is
clear from the enhanced flux in their wavelength range.
\nocite{pmh+99}\nocite{ljs01}

We have fitted the observed spectrum with the models discussed by previous
authors, i.e.\ black body, LTE atmosphere and NLTE atmosphere
spectra. 
Following earlier authors, we also added a thermal plasma in 
collisional ionization equilibrium to a NLTE atmosphere.
In Fig.\,\ref{spectrum} we show the spectra of 
the best fit black body and the combination of the atmosphere and the
plasma.
The model parameters are given in Table\,\ref{tabmod}.

The interstellar absorption edges due to Ne\,I (at $\lambda$\,14.30\,\AA)
and Fe-L (at $\lambda$\,17.54\,\AA) are recognizable in the models and in the
data, and confirm that interstellar absorption is appreciable.
It is clear that neither model is acceptable.
The observed spectrum shows much more structure than a black body;
but not the structure expected for a calculated atmospheric spectrum.
This is true for a gravitational acceleration of 10$^9$\,cm\,s$^{-2}$ and 
effective temperatures between 8.00-9.15\,10$^5$\,K.

The combination of a thermal plasma with the NLTE atmosphere indeed
adds lines, but the strengths of these are well below those of the observed
emission complexes between $\lambda$\,14-16\,\AA. The flux contribution of 
the plasma is
so small that the parameters of the atmospheric model are virtually
the same for both models with and without an added thermal plasma.
In some calculations we allow an arbitrary global shift in the wavelengths 
of the lines from the thermal plasma to take into account the
current accuracy of the HETG-MEG absolute wavelength calibration;
this doesn't change any of the conclusions. 
Similarly, varying the abundances of iron or oxygen in the thermal 
plasma only results in marginal improvement; only some emission lines 
can be reproduced better this way.
Assuming different interstellar absorption densities for the two model
components, complicates the analysis further with no better results.

If one tries to enhance the emission line strengths by changing
the temperature or the emission measure of the plasma, one
always finds very strong lines at wavelengths where none are observed.
For example, whenever the strength of the Fe\,XVII lines 
at $\lambda$\,15.014 and 15.265\,\AA\ are
enhanced to bring their fluxes to the observed level, the lines from
the same ion at $\lambda$\,12.124 and 12.264\,\AA\ become too strong.
It also produces a thermal continuum which is 
significantly higher between $\lambda$\,7-12\,\AA\ than the observed flux.

\begin{table}
\begin{tabular}{|p{2.5cm}|p{2.5cm}|p{2.5cm}|}
\hline
Theoretical 	&Fitted 	&Relative flux\\
wavelength (\AA)&wavelength (\AA)	&\\
\hline
15.014			&15.05 {\tiny $\pm$\,0.01}	&0.63 {\tiny -0.22\,/\,+0.25}\\
15.265			&15.276 {\tiny $\pm$\,0.007}	&1.00 {\tiny -0.24\,/\,+0.27}\\
16.780			&16.79 {\tiny $\pm$\,0.01}	&1.61 {\tiny -0.78\,/\,+1.40}\\
17.055\,/\,17.100		&17.10 {\tiny $\pm$\,0.01}	&3.45 {\tiny -1.44\,/\,+2.20}\\
\hline
\end{tabular}
\caption{Parameter values for the fit of the identified Fe\,XVII lines with
gaussian profiles. The two lines at $\lambda$\,17.055 and 17.100\,\AA\ are 
fitted with one
profile. The fluxes are printed relative to the flux of the line at 
$\lambda$\,15.265\,\AA,
and have been corrected for interstellar absorption with an absorption measure
of 0.99\,10$^{22}$\,cm$^{-2}$.
Note that this emission feature may have a contribution of
the O\,VIII Lyman\,$\gamma$ line.}
\label{gauss}
\end{table}

To compare the relative strengths of the identified Fe\,XVII lines with
theoretical predictions, we have fitted them with Gaussian profiles,
as printed in Table\,\ref{gauss}.
No thermal plasma in collisonal ionization equilibrium can reproduce the 
observed
relative strengths of the Fe\,XVII lines, even when we allow a continuous
distribution of emission measures as a function of the temperature
(the differential emission measure model, see Liedahl et al.\ 2001).
Also the strength of the Fe\,XVIII lines between $\lambda$\,14.3-14.6\,\AA\ 
can't be explained by such plasma. These results remain true when the 
abundances are varied from solar, because they affect the relative strenghts 
of lines from different elements, not those from the same element.

The main problem is the ratio of the 15.014\,/\,15.265 lines, for which
optically thin models predict a value of 3-4 for any density or
temperature, while we observe 0.63. The ratio of the 16.780 and 
17.055\,/\,17.100 lines to the 15.265 line is however consistent with an
optically thin plasma. Dropping the optically thin assumption, the
15.014 line can be brought into agreement with the 16.780 and 
17.055\,/\,17.100 fluxes for pure Fe\,XVII column densities of the
order of 10$^{19}$\,cm$^{-2}$, but also in that case the predicted
15.014\,/\,15.265 ratio still remains similar as in the
optically thin case. Thus, also for the optically
thick model the 15.265 line is 6 times weaker than observed.

We conclude that none of the earlier models describe the observed
X-ray spectrum of \rx.

We investigate whether the spectrum
shows signatures of a photo-ionized plasma, as listed by Liedahl et
al.\ (2001), viz.\ a) narrow radiative recombination continua; b) large ratio
$(x+y+z)/w$ of the He-like triplet; c) the weakness of L-shell iron
emission compared to K-shell emission from lighter elements.
The strength of the Fe\,XVII lines and the relative weakness of the
Ne\,IX lines argues against a photo-ionized plasma. The observed spectrum 
also does not show evidence for narrow recombination continua.
The strengths of the individual Ne lines in the spectrum are too small to
determine their ratios. 
We conclude that a simple photo-ionized plasma also doesn't 
describe the observations.

\section{P\,Cygni profiles indicate a stellar wind}

\begin{figure}[!t]
\epsfig{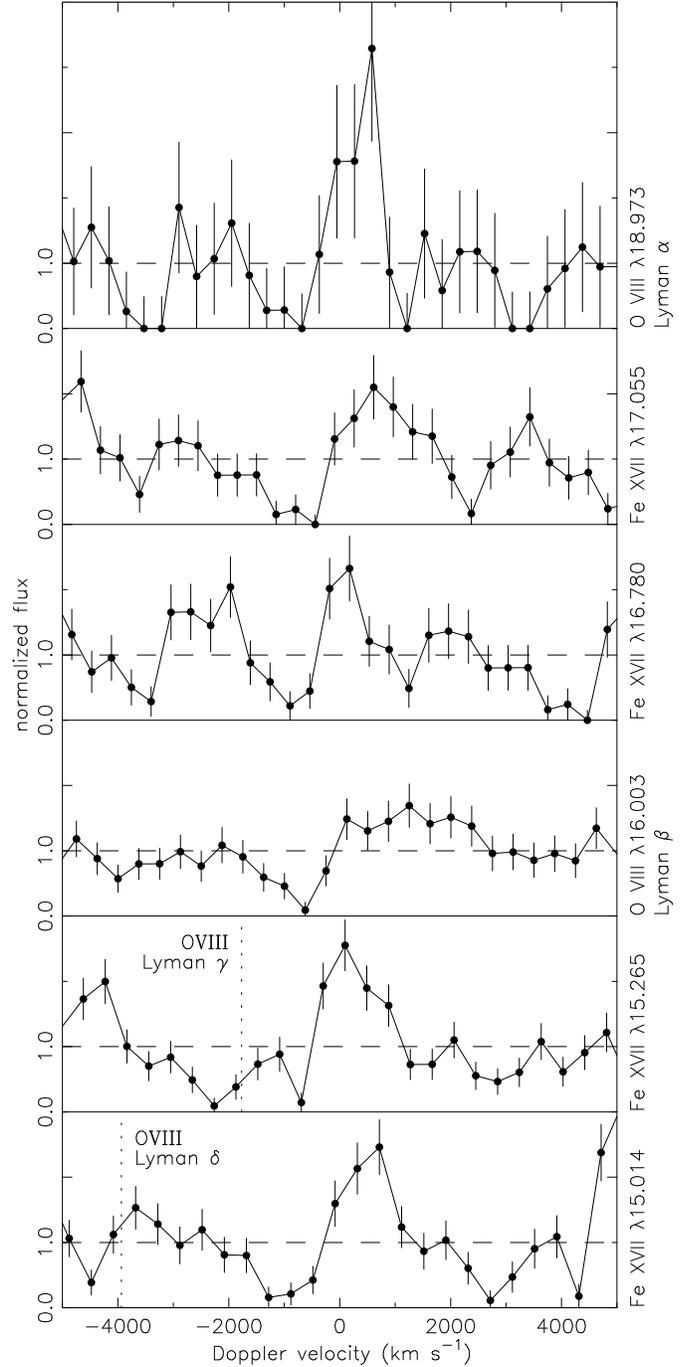}
\caption{Profiles of some strong, individually identifiable lines
in the spectrum of \rx. The average flux in each frame has been
normalized to unity; this level is indicated with the dashed lines.
The wavelength scale is given in Doppler velocities, based on central
wavelengths shown on the right of each frame; the wavelength resolution
is the instrumental resolution, i.e.\ more points are shown than
in Fig.\,\ref{spectrum}.
The central wavelengths of the oxygen Lyman $\gamma$ and $\delta$ lines 
are indicated with vertically dotted lines in the two lower frames.}
\label{pcygni}
\end{figure}

Considering that none of the previously suggested models 
describes the data, we have another look at the observed
spectrum.
Fig.\,\ref{pcygni} gives an expanded view of the five
strong, individually identifiable lines and the spectrum
around the O\,VIII Lyman\,$\beta$ line at 16.003\,\AA.
The similarity between the profiles of these lines
indicates that the P\,Cygni structure seen in them is real.
One would expect to see a similar structure for the
Lyman\,$\gamma$ and $\delta$ lines; inspection of the
spectra near their wavelengths (15.175 and 14.817\,\AA, 
respectively) indeed shows indication of redward emission
and blueward absorption. 
We conclude from the P\,Cygni profiles 
that the stronger individually identifiable lines 
(see Sect.\,5) in our spectrum arise in a wind.

The absorption parts of the O\,VIII Lyman\,$\alpha$ and $\beta$ 
and Fe\,XVII $\lambda$\,17.055\,\AA\ lines are saturated;
the absorption part of the O\,VIII Lyman\,$\delta$ line is not
saturated. For the remaining lines the situation is ambiguous.

Strong P\,Cygni lines are present in the spectra of high-mass
stars, e.g. Wolf-Rayet stars, and also in the spectra of
systems with an accretion disk. 
The radial velocity curve and the absence of strong stellar lines 
in the optical spectrum, exclude
the presence of a high-mass star in \rx\ (see Sect.\,2).
Thus, the P\,Cygni profiles in the X-ray spectrum of
\rx\ are more likely related to the presence of an
accretion disk.

In cataclysmic variables, the terminal wind velocities
are often fairly high, reaching values of 5000\,km\,s$^{-1}$
(e.g.\ Woods et al.\ 1992). The similar velocity of the
transient jet seen in the optical has been used by Motch 
(1998) to argue that the accreting object is a white dwarf.
Remarkably, the highest absorption velocities seen in
Fig.\,\ref{pcygni} are rather lower, of the order of 1500\,km\,s$^{-1}$.
\nocite{wvc+92}

It seems possible that a combination of an atmosphere spectrum
with P\,Cygni profiles from a wind gives a good description
of the X-ray spectrum of \rx. It is beyond the scope of this article
to construct such a model. At this point we repeat our earlier
statement that no obvious atmospheric features can be seen in
the X-ray spectrum, as shown in Fig.\,\ref{spectrum}.

\section{Discussion}
The observed X-ray spectrum of \rx\ is dominated by emission
features, and is not the continuum with absorption features
expected on the basis of stellar atmosphere models,
nor is it an optically thin spectrum of a thermal plasma 
in collisional ionization equilibrium, or a photo-ionized plasma.
Instead, we suggest that the spectrum is a combination of a stellar
atmosphere and a wind.

The bulk of the luminosity of the models fitted
by earlier authors lies outside the observed X-ray regimes, 
at softer energies.
As a result the luminosities derived from the same observations
range from above the Eddington limit for a black
body to less than 0.001 of the Eddington limit
for the spectrum of a NLTE atmosphere, as can be
observed in Table\,\ref{tabmod}.
The uncertain magnitude of the interstellar
absorption adds to the uncertainty.
This also may be seen in Table\,\ref{tabmod}, by
comparing the same model fitted to data of different satellites.

Therefore we cannot predict the bolometric
luminosity corresponding to the X-ray spectrum of \rx.
This remains true for models with a wind.
We can only determine the luminosity in the
observed range of 8-20\,\AA\ at $\sim\,10^{35}\,\ergs$ for an
assumed distance of 1\,kpc.
As a result, we can no longer be certain that
\rx\ is as luminous as expected for a white
dwarf steadily burning helium at its surface.

One of the other supersoft sources, Cal\,83,
has been observed with the reflection grating
spectrograph of XMM, with a resolution  of
0.05\,\AA\ in the range of interest for this
softer source, 20-38\,\AA\ (Paerels et al.\ 2001).
At first glance, the observed spectrum of Cal\,83
could consist of either emission features with
a width of $\sim$\,1\,\AA, or a continuum  with 
many absorption features of comparable width.
Paerels et al.\ prefer the latter interpretation,
but cannot match the absorption features with features
from an atmosphere dominated by CNO, or from a model
with element abundances as appropriate for the LMC,
up to Fe. 
At the moment then, the bolometric flux of Cal\,83 also
is unknown.
\nocite{prh+01}

Before definite conclusions can be drawn about the nature of the
supersoft sources, more progress is required in the theoretical
description of their X-ray spectra. In particular, it seems necessary
to develop models that combine atmosphere models with winds. 
The atmosphere models themselves can be improved with respect to the current
models in several ways. More elements and ions can be included
in the computation of the atmospheric structure, with a more 
complete account of the line-blanketing effects.
Gradients in temperature and gravitational acceleration, as can be expected
for accretion disk spectra, cannot be ruled out.
Non-static or outflowing atmospheres may have to be considered.
The question arises whether the jet observed
in \rx\ and the winds observed in several supersoft sources
contribute to their X-ray spectra.

\bigskip\noindent
\emph{Acknowledgements}\\
We thank Herman Marshall at the Chandra Science Center
for his support in this observation.

The Laboratory for Space Research Utrecht is supported
financially by NWO, the Netherlands Organization for Scientific
Research.

\end{document}